\documentclass[12pt,psfig]{article}
\usepackage[centertags]{amsmath}
\usepackage{amssymb}
\newcommand{\Z}{{\mathbb Z}}
\newcommand{\R}{{\mathbb R}}
\begin{document}
\begin{flushright}
\baselineskip=12pt
{ACT-04-05}\\
{MIFP-05-17}\\
\end{flushright}
\def\IZ{Z\kern-.5em Z}

\begin{center}
{\LARGE \bf A K-Theory Anomaly Free Supersymmetric Flipped SU(5) Model 
from Intersecting Branes.\\}
\vglue 1.00cm
{C.-M. Chen$^{\spadesuit}$ \footnote
{cchen@physics.tamu.edu}, G. V. Kraniotis$^{\spadesuit}$ \footnote
 {kraniotis@physics.tamu.edu} , V. E. Mayes$^{\spadesuit}$ \footnote{eric@physics.tamu.edu},\\ D. V. Nanopoulos $^{\spadesuit,\clubsuit,\diamondsuit}$\footnote{dimitri@physics.tamu.edu}, and J. W. Walker $^{\spadesuit}$ \footnote{jwalker@physics.tamu.edu}\\}
\vglue 0.2cm
    {$\spadesuit$ \it  George P. and Cynthia W. Mitchell Institute for
Fundamental Physics,\\
 Texas A$\&$M University, College Station TX,
77843, USA  \\}
{$\clubsuit$ \it Astroparticle Physics Group, Houston Advanced Research Center (HARC)\\
Mitchell Campus, Woodlands, TX 77381, USA\\}
{$\diamondsuit$ \it  Academy of Athens, Division of Natural Sciences,  \\}
{\it 28 Panepistimiou Avenue, Athens 10679, Greece }
\baselineskip=12pt

\vglue 2.5cm
ABSTRACT
\end{center}

We construct an $N=1$ supersymmetric three-family flipped $SU(5)$ model from
type IIA orientifolds on $ T^6/(\Z_2\times \Z_2)$ with D6-branes
intersecting at general angles. 
The model is constrained by the requirement that Ramond-Ramond tadpoles 
cancel, the supersymmetry conditions, and that the gauge boson coupled to the 
$U(1)_X$ factor does not get a string-scale mass via a generalised 
Green-Schwarz mechanism. 
The model is further constrained by requiring cancellation of K-theory charges.
The spectrum contains a complete grand unified
and electroweak Higgs sector, however the latter in a non-minimal number of 
copies. In addition, it contains extra  matter
both in bi-fundamental and vector-like representations as well as two copies of
matter in the symmetric representation of $SU(5)$.

{\rightskip=3pc
\leftskip=3pc
\noindent
\baselineskip=20pt

}

\vfill\eject
\setcounter{page}{1}
\pagestyle{plain}
\baselineskip=14pt

\section{Introduction}

One of the main goals of string phenomenology is to derive standard model 
physics in a convincing way and to embed the latter into a unified 
description of gravitational and gauge forces.

In the intersecting D-branes approach in type II string theory
\cite{BDL,BACHAS,BERLIN,AFIRU1}, gravity is mediated in the entire 
10-dimensional bulk by the exchange of closed strings. On the other hand, 
the gauge and matter fields are localized respectively on the D-brane,
and at pairwise D-brane intersections, and 
correspond to open string excitations. One of the outstanding issues 
is the breaking of supersymmetry. There are at least two scenarios of how 
supersymmetry breaking can be realized: In the first case, the various 
open string sectors 
break supersymmetry while the closed string sector may preserve supersymmetry.
In order to solve the hierarchy problem this scenario normally requires the 
existence of large extra dimensions, transverse to the D-branes of the standard model \cite{CIM,BKL1,HON}.  In the second scenario, all open string sectors preserve $N=1$ supersymmetry, and all D-branes together with the orientifold planes, are mutually supersymmetric. Then the closed string sector breaks 
supersymmetry, which manifests itself as soft supersymmetry-breaking terms in the effective 
action of the open string matter fields. The spontaneous supersymmetry breaking can be achieved by internal background fluxes of 
closed string field strengths $<G_{ijk}>\not = 0$ \cite{Dieter,Cvetic,cjan05}.

In this latter case, one in general has to introduce in addition to D6-branes
orientifold O6-planes, which can be regarded as branes of negative
RR-charge and tension. For a general Calabi-Yau compact space
these orientifold planes wrap special Lagrangian 3-cycles
calibrated with respect to the real part of the holomorphic 3-form
$\Omega_3$ of the Calabi-Yau compact space. In order to preserve $N=1$ supersymmetry all the D6-branes must be calibrated by the 3-form $\Omega_3$. Calibrated 
geometries lead to volume mininization in homology. Adopting this 
philosophy for solving the gauge hierarchy problem 
and  other instabilities of non-supersymmetric models 
from intersecting branes, we recently constructed a 
three-generation $N=1$ supersymmetric flipped $SU(5)$ model \cite{nkwmm}, whose gauge symmetry included $SU(5)\times U(1)_X$ symmetry, from type IIA orientifolds on $ T^6/(\Z_2\times \Z_2)$ with intersecting D6-branes. For other examples of Grand Unified Models 
as well as for a  complete set of references in the field, the reader should consult 
\cite{MIRIAM,REVIEWS1}.

The flipped $SU(5)\times U(1)_X$ model \cite{BARR,IJDJ} is a well 
motivated example of a Grand Unified Theory (GUT), and  
had been extensively studied in the closed string era of the
heterotic compactifications \cite{ANTON,YUAN}. From the
theoretical point of view this motivation was coming from the fact
that its symmetry breaking requires only ${\bf 10}$ and $\bf
{\overline{10}}$ Higgs representations at the grand unification scale,
as well as ${\bf 5}$ and ${\bf \bar{5}}$ Higgs representations at the
electroweak scale, which were consistent with the
representations of $SU(5)$ allowed by the unitarity condition with
gauge group at level 1 \cite{PETER,HERBI} \setcounter{footnote}{0}\footnote{Thus attempts
to embed conventional grand unified theory (GUT) groups such as
$SU(5)$ or $SO(10)$ in heterotic string required more complicated
compactifications, but none of these has been completely
satisfactory. Constructions with the minimal option to embed just
the standard model gauge group, were plagued with at least extra
unwanted U(1) factors.}. From the phenomenological point of view
flipped $SU(5)\times U(1)_X$ \cite{BARR,IJDJ} has a number of
attractive features in its own right \cite{Dimitri}. For example,
it has a very elegant missing-partner mechanism for suppressing
proton decay via dimension-5 operators \cite{IJDJ}, and is
probably the simplest GUT to survive experimental limits on proton
decay \cite{JDJW}. These considerations motivated the derivation
of a number of flipped $SU(5)$ models from constructions using
fermions on the world sheet \cite{ANTON,YUAN}.
Also, non-supersymmetric flipped $SU(5)$ models have been 
produced in \cite{PDJ} using D6-branes wrapping toroidal 3-cycles and  
when the wrapping space is the $T^6/\Z_3$ orbifold.

The wrapping numbers of the various stacks in \cite{nkwmm} were constrained 
by the
requirement of RR-tadpole cancellation as well as the
supersymmetry conditions. Tadpole cancellation ensures the absence
of non-abelian anomalies in the emergent low-energy quantum field
theory.  A generalised Green-Schwarz mechanism ensures that 
the gauge bosons associated with all anomalous $U(1)$s acquire
string-scale masses \cite{IRU}, but the gauge bosons of some
non-anomalous $U(1)$s can also acquire string-scale masses
\cite{IMR}; in all such cases the $U(1)$ group survives as a
global symmetry.  Thus we had also to ensure the flipped  $U(1)_X$
group remained a gauge symmetry by requiring that its gauge boson
did {\it not} get such a mass. The gauge symmetry of the model included a 
$USp(2)\cong SU(2)$ factor associated with the presence of filler branes (D-branes that wrap 
3-cycles which are invariant under the orientifold action).  The low energy spectrum of the model we constructed was free from any 
$SU(2)$ global gauge anomalies (associated with the fourth non-trivial homotopy 
group of $SU(2)$: $\pi^4(SU(2))=\Z_2$) since the number of the corresponding 
fermion doublets was even \cite{Witten} \footnote{This was the case since the number of intersections of 
the filler branes to the other stacks of D6-branes was even.}.
The model however, suffered from a number of serious phenomenological drawbacks. Among them, the global $U(1)$ symmetries of the model, that arise after the G-S anomaly
cancellation mechanism, did forbid some of the Yukawa couplings
required for mass generation, as well as the couplings responsible 
for the elegant solution of the  doublet-triplet splitting problem in
flipped $SU(5)$ \cite{nkwmm}. The model also included a lot of exotic matter both in 
bi-fundamental and vector-like representations, as
well as two copies of matter in the symmetric representation of
$SU(5)$. Furthermore, three adjoint (24-plets) $(N=1)$ chiral multiplets were
also present. 

It has been argued by a number of authors \cite{MM,EW,UR,SCM}, that for the D-branes to consistently wrap the 3-cycles of the compact space,  additional conditions on their 
wrapping numbers have to be satisfied beyond those described above, which stem from the K-theory interpretation of D-branes. In particular, it has been argued that often it is K-theory which fully classifies the RR-charges of D-branes and 
not the  ordinary homology theory \footnote{For some computations 
of the K-theory groups, that make use of Bott's periodicity 
theorem and Atiyah's real K-theory \cite{Atiyah},  for D-branes on top of orientifold planes see \cite{Gukov}.}. 
This approach was also motivated by the work in \cite{SEN} in which the non-BPS D-branes were constructed as bound states of brane-anti-brane pairs. These 
constructions were interpreted in terms of K-theory in \cite{EW}.
The model constructed in \cite{nkwmm} did not 
satisfy all the additional constraints from K-theory derived in \cite{SCM} 
for the particular orientifold background.

It is the purpose of this Letter, to incorporate these additional constraints,
which ensure that discrete K-theory charge cancels, 
into the search for building a viable flipped $SU(5)$ model.
Interestingly, as we shall see in the main body of the paper, although 
the new model 
constructed, which is compatible with the above consistency requirements, 
does contains a lot of extra matter, it does not contain exotic matter 
other then  two 
copies in the symmetric ($\overline{\bf 15}$) representation of 
$SU(5)$ in the spectrum on  the sector of D6-branes at generic angles (i.e.
those that wrap 3-cycles not invariant under the anti-holomorphic involution).
However, the model still possesses exotic matter  from the intersections of 
filler branes with the other stacks of D6-branes, which is charged under 
both the flipped $SU(5)$ and filler branes gauge symmetries.

The material of this Letter is organized as follows. In section 2, we provide 
the consistency as  well as the other conditions described above
for the $T^6/(\Z_2\times \Z_2)$ orientifold 
which are used for the construction of our model. In section 3, after a 
short introduction to basic flipped $SU(5)$ phenomenology we present the 
model, which is a consistent solution of  all the constraints we described 
above including those from K-theory. Finally, section 4 is used for our conclusions.

\section{Definitions and Conditions for Intersecting Brane Models on a $\mathbf{T^6/(\Z_2\times \Z_2)}$ Orientifold}
\subsection{Basic Configuration}

Consider type IIA theory on the $\mathbf{T^6/(\Z_2\times \Z_2)}$
orientifold, where the orbifold group $\Z_2\times \Z_2$ generators
$\theta$, $\omega$ act on the complex coordinates $(z_1,z_2,z_3)$
of $\mathbf{T^6}$ as
\begin{eqnarray}
\theta:(z_1,z_2,z_3)\rightarrow(-z_1,-z_2,z_3) \nonumber \\
\omega:(z_1,z_2,z_3)\rightarrow(z_1,-z_2,-z_3)
\end{eqnarray}
This $T^6/(\Z_2\times\Z_2)$ structure was first introduced in \cite{Cvetic} 
and further studied in \cite{MIRIAM} \footnote{See also \cite{LEIGH}.},
and we will use the same notations here.
We implement an orientifold projection $\Omega R$, where $\Omega$
is the world-sheet parity, and $R$ acts as
\begin{equation}
R:(z_1,z_2,z_3)\rightarrow(\overline{z}_1,\overline{z}_2,\overline{z}_3)
\end{equation}

With the wrapping numbers $(n^i, m^i)$ along the canonical basis
of homology one-cycles $[a_i]$ and $[b_i]$ the complete cycle on a
$\mathbf{T^2}$ is given by $n^i[a_i]+m^i[b_i]$.  Note that a
tilted complex structure is allowed by setting
$[a'_i]\equiv[a_i]+\frac{1}{2}[b_i]$ so we rewrite the expression
of the one-cycle by $n^i[a'_i]+ 2^{-\beta_i} l^i[b_i]$ where
$l^i=m^i$, $\beta_i=0$ if the $i$th torus is not tilted and $l^i=2m^i+n^i$,
$\beta_i=1$ if it is tilted.

Therefore the homology three cycle $[\Pi_a]$ for a stack a of
D6-brane and its orientifold image $[\Pi_a']$ can be written as
\begin{equation}
[\Pi_a]=\prod_{i=1}^{3}(n^i_a[a_i]+2^{-\beta_i}l^i_a[b_i]),\;\;\;
[\Pi_{a'}]=\prod_{i=1}^{3}(n^i_a[a_i]-2^{-\beta_i}l^i_a[b_i])
\end{equation}
and the O6-plane associated with the four orientifold projections
$\Omega R$, $\Omega R\theta$, $\Omega R \omega$, and $\Omega
R\theta\omega$ is
\begin{eqnarray}
[\Pi_{O6}]&=&[\Pi_{\Omega R}]+[\Pi_{\Omega R \omega}]+[\Pi_{\Omega
R \theta\omega}]+[\Pi_{\Omega R \theta}] \nonumber \\ &=&
2^3[a_1][a_2][a_3] -2^{3-\beta_2-\beta_3}[a_1][b_2][b_3] \nonumber
\\ & & -2^{3-\beta_1-\beta_3}[b_1][a_2][b_3]
-2^{3-\beta_1-\beta_2}[b_1][b_2][a_3]
\end{eqnarray}
It is  convenient  for model building purposes to use a set of parameters introduced in \cite{MIRIAM}
\begin{eqnarray}
& A_a=-n^1_a n^2_a n^3_a,\; B_a=n^1_a l^2_a l^3_a,\; C_a=l^1_a
n^2_a
l^3_a,\; D_a=l^1_a l^2_a n^3_a  \nonumber \\
& \tilde{A_a}=-l^1_a l^2_a l^3_a,\; \tilde{B_a}=l^1_a n^2_a
n^3_a,\; \tilde{C_a}=n^1_a l^2_a n^3_a,\; \tilde{D_a}=n^1_a n^2_a
l^3_a \label{wrapparameter}
\end{eqnarray}

\subsection{The Spectrum}

Chiral matter particles are formed from open strings with two ends
attaching on different stacks.  The multiplicity (${\cal M}$) of
the corresponding bi-fundamental representation is given by the
intesection numbers ($I_{ab}$) between different stacks of branes by using
Gra$\ss$mann algebra $[a_i][b_j]=-[b_j][a_i]=\delta_{ij}$ and
$[a_i][a_j]=-[b_j][b_i]=0$.  It should be noted that the initial $U(N_a)$ gauge
group supported by a stack of $N_a$ identical D6-branes is broken
down by the $\Z_2\times \Z_2$ symmetry to a subgroup $U(N_a/2)$
\cite{Cvetic}.  In Table 1 we exhibit the 
generic chiral spectrum for the $T^6/(\Z_2\times\Z_2)$ orientifold
for D6-branes at generic angles \cite{MIRIAM}.  
A zero intersection number between two branes implies that the branes 
are parallel on at least one torus.
At such kind of intersection additional
non-chiral (vector-like) multiplet pairs from $ab+ba$, $ab'+b'a$,
and $aa'+a'a$ sectors can arise \cite{PAUL}.
The multiplicity of these non-chiral multiplet pairs is given by
the remainder of the intersection product, neglecting the null
sector. For example, if $(n^1_a l^1_b - n^1_b l^1_a)=0$ in $
I_{ab}=[\Pi_a][\Pi_b]=2^{-k}\prod_{i=1}^3(n^i_a l^i_b - n^i_b
l^i_a) $,
\begin{equation}
{\cal M}\left[\left(\frac{N_a}{2},\frac{\overline{N_b}}{2}\right)
+\left(\frac{\overline{N_a}}{2},\frac{N_b}{2}\right)\right]
=\prod_{i=2}^3(n^i_a l^i_b - n^i_b l^i_a)
\end{equation}
where we have assumed that, $k\equiv \beta_1+\beta_2+\beta_3=0$, i.e. all tori are 
untilted. 

Strings stretching between a brane in stack $a$ and its mirror
image $a'$ yield chiral matter in the antisymmetric and symmetric
representations of the group $U(N_a/2)$ with multiplicities
\begin{equation}
{\cal M}(({\rm A}_a)_L)=\frac{1}{2}I_{aO6},\;\; {\cal M}(({\rm
A}_a+{\rm S}_a)_L)=\frac{1}{2}(I_{aa'}-\frac{1}{2}I_{aO6})
\end{equation}
so that the net total of antisymmetric and symmetric
representations are those in Table 1. Also 
\begin{equation}
I_{aa'}=[\Pi_a][\Pi_{a'}]=-2^{3-k}\prod_{i=1}^3 \,n^i_a l^i_a
\end{equation}
\begin{equation}
I_{aO6}=[\Pi_a][\Pi_{O6}]=2^{3-k}(\tilde{A_a}+\tilde{B_a}+\tilde{C_a}+\tilde{D_a})
\end{equation}
This distinction is critical, as we require independent use of the
paired multiplets such as $(\mathbf{10},\mathbf{\overline{10}})$
which are masked in the corresponding expressions given in Table 1. 

\begin{table}[h]
\renewcommand{\arraystretch}{1.5}
\center
\begin{tabular}{|c||c|}
\hline

Sector & Representation   \\ \hline \hline

$aa$ & $U(N_a /2)$ vector multiplet and 3 adjoint chiral multiplets \\
\hline

$ab+ba$ & $ {\cal M}(\frac{N_a}{2},
\frac{\overline{N_b}}{2})=I_{ab}=2^{-k}\prod_{i=1}^3(n^i_a l^i_b -
n^i_b l^i_a) $  \\ \hline

$ab'+b'a$ & $ {\cal M}(\frac{N_a}{2},
\frac{N_b}{2})=I_{ab'}=-2^{-k}\prod_{i=1}^3(n^i_a l^i_b + n^i_b
l^i_a) $
\\ \hline

$aa'+a'a$ & $  {\cal M}({\rm
Anti}_a)=\frac{1}{2}(I_{aa'}+\frac{1}{2}I_{aO6}) =
-2^{1-k}[(2A_a-1)\tilde{A_a}-\tilde{B_a}-\tilde{C_a}-\tilde{D_a}]
 $ \\
 & $ {\cal M}({\rm
Sym}_a)=\frac{1}{2}(I_{aa'}-\frac{1}{2}I_{aO6}) =
-2^{1-k}[(2A_a+1)\tilde{A_a}+\tilde{B_a}+\tilde{C_a}+\tilde{D_a}]
$   \\ \hline

\end{tabular}
\caption{General chiral spectrum on D6-branes at generic angles, for 
$T^6/(\Z_2\times \Z_2)$ orientifold.}
\end{table}

\subsection{Consistency and Supersymmetry Conditions}

The Ramond-Ramond tadpole cancellation needs to be satisfied by
requiring the total homology cycle charge of D6-branes and
O6-planes to vanish, namely we require
\begin{equation}
\sum_a N_a[\Pi_a]+\sum_a N_a[\Pi_{a'}]-4[\Pi_{O6}]=0
\end{equation}
By introducing the filler branes wrapping cycles along the 
four O6-planes, we can
rewrite the above equation in terms of the parameters defined in (\ref{wrapparameter})  as
\begin{eqnarray}
&&-2^k N^{(\Omega R)}+\sum_a N_a A_a=-2^k N^{(\Omega R \omega)}+\sum_a N_a B_a = \nonumber \\
&&-2^k N^{(\Omega R \theta\omega)}+\sum_a N_a C_a=-2^k N^{(\Omega
R \theta)}+\sum_a N_a D_a = -16 \label{RR}
\end{eqnarray}

Although the total non-Abelian anomaly cancels automatically when
the RR-tadpole conditions are satisfied, additional mixed
anomalies like the mixed gravitational anomalies which generate
massive fields are not trivially zero \cite{Cvetic}.  These anomalies are
cancelled by a generalized Green-Schwarz (G-S) mechanism which
involves untwisted Ramond-Ramond forms.  The couplings of the four
untwisted Ramond-Ramond forms $B^i_2$ to the $U(1)$ field strength
$F_a$ of each stack $a$ are
\begin{eqnarray}
 N_a \tilde{B_a} \int_{M4}B^1_2\wedge \textrm{tr}F_a,  \;\;
 N_a \tilde{C_a} \int_{M4}B^2_2\wedge \textrm{tr}F_a
  \nonumber \\
 N_a \tilde{D_a} \int_{M4}B^3_2\wedge \textrm{tr}F_a,  \;\;
 N_a \tilde{A_a} \int_{M4}B^4_2\wedge \textrm{tr}F_a
\end{eqnarray}
These couplings determine the linear combinations of $U(1)$ gauge
bosons that acquire string scale masses via the G-S mechanism. In
flipped $SU(5)\times U(1)_X$,  the symmetry $U(1)_X$ must remain an unbroken 
gauge symmetry  so that it may remix to help generate the standard
model hypercharge after the breaking of $SU(5)$.  Therefore, we
must ensure that the gauge boson of the flipped $U(1)_X$ group
does not receive such a mass. The $U(1)_X$ is a linear combination
(to be identified in section 3.2) of the $U(1)$s from each stack :
\begin{equation}
U(1)_X=\sum_a c_a U(1)_a
\end{equation}
The corresponding field strength must be orthogonal to those that
acquire G-S mass.  Thus we demand :
\begin{eqnarray}
\sum_a c_a N_a \tilde{B_a} =0, \;\; \sum_a c_a N_a \tilde{C_a} =0
  \nonumber \\
\sum_a c_a N_a \tilde{D_a} =0, \;\; \sum_a c_a N_a \tilde{A_a} =0
\label{GSeq}
\end{eqnarray}

The condition to preserve $\emph{N}=1$ supersymmetry in four
dimensions is that the rotation angle of any D-brane with respect 
to the orientifold plane is an element of $SU(3)$ \cite{
BDL,Cvetic}. Considering the angles between each brane and the R-invariant
axis of $i^{\mathrm{th}}$ torus $\theta^i_a$, we
require
$\theta^1_a + \theta^2_a + \theta^3_a = 0$ mod $2\pi$.  
This means $\sin(\theta^1_a
+
  \theta^2_a + \theta^3_a)= 0$ and $\cos(\theta^1_a +
  \theta^2_a + \theta^3_a)= 1 > 0 $.
We define 
\begin{equation}
\tan\theta^i_a=\frac{2^{-\beta_i}l^i_a R^i_2}{n^i_a R^i_1}
\end{equation}
where $R^i_2$ and $R^i_1$ are the radii of the $i^{\mathrm{th}}$
torus. The above supersymmetry conditions
 can be recast in terms of the parameters
defined in (\ref{wrapparameter}) as follows \cite{MIRIAM}:
\begin{eqnarray}
x_A\tilde{A_a}+x_B\tilde{B_a}+x_C\tilde{C_a}+x_D\tilde{D_a}=0
\nonumber \\
A_a/x_A + B_a/x_B + C_a/x_C + D_a/x_D <0
\end{eqnarray}
where $x_A$, $x_B$, $x_C$, $x_D$ are complex structure parameters
composed of $R^i_2$ and $R^i_1$, all of which share the same sign 
\cite{MIRIAM}. In what follows we consider the case $k=0$.

\subsection{The K-Theory Constraints}


In the previous section, the consistency conditions for having a model free of
RR tadpoles were stated.  These conditions
essentially translate into constraints on the allowed homology cycles.  
However, as we discussed in the introduction it has been argued that it 
is K-theory which fully classifies the RR-charges of D-branes and not 
the ordinary homology theory \cite{MM,EW,UR,SCM,MC}.
Because of this, there are
additional consistency constraints related to cancellation of the K-theory
charges \cite{UR,SCM}.  These additional constraints are not
visible through
homology.  In this Letter, we improve upon our earlier work \cite{nkwmm} with a
model that satisfies these K-theory constraints as
well as all the other conditions for the $T^6/(\Z_2\times\Z_2)$ orientifold described 
in section 2.3.

In type I superstring theory there exist non-BPS D-branes carrying
non-trivial K-theory $\mathbf{\Z_2}$ charges.  To avoid this
anomaly it is required that in compact spaces these non-BPS branes
must exist in an even number \cite{UR}.  
If we consider a type I non-BPS D7-brane
($\widehat{\textrm{D7}}$-brane),  we may regard it as a pair
of D7-brane and its world-sheet parity image
$\overline{\textrm{D7}}$-brane in type IIB theory, i.e. 
$\widehat{\textrm{D7}}={\rm D7}+\overline{\rm D7}/\Omega$ .  There are
three different kinds of non-BPS ($\widehat{\textrm{D7}}$)-branes,
denoted as $\widehat{\textrm{D7}_i}$, where $i=1,2,3$ labels the two-torus 
where the $\widehat{\textrm{D7}}$ does not wrap. By construction there are 
three pairs D$7_i$, $\overline{{\rm D7}}_i$ in type IIB theory \cite{SCM}.
These D7-brane pairs, as well as other D-brane pairs in type IIB
theory, can be explicitly expressed by the homology 3-cycles in 
type IIA theory 
as listed in Table 2 \footnote{In type IIB picture D$5_i$ stands for a 
D5-brane wrapping the $i^{th}$ two torus.}.

It is reasonable to take the branes in Table 2 as a basis of a
magnetized model (obviously they are in terms of the homology
one-cycles). We can see that a general D6-brane three-cycle in
type IIA theory  is composed of these brane pairs, i.e., a general D6-brane is a linear combination of these brane
pairs, which is why we should take the K-theory constraints into account since
the numbers of the pairs given by wrapping numbers are not
trivially even.

We do not have to worry about the K-theory charge
contributed by D5 and D9-branes since the RR-tadpole conditions in
(\ref{RR}) guarantee the even numbers if we choose the number of
the filler branes to be even, which is not difficult to achieve.  The
real problem comes from D3 and D7-branes, though they do not
contribute to the standard RR charges.  The K-theory conditions
for a $\mathbf{\Z_2\times \Z_2}$ orientifold were derived in
\cite{SCM} and are given by
\begin{eqnarray}
\sum_a N_a l^1_a l^2_a l^3_a = \sum_a N_a \tilde{A}_a = 0 \textrm{
mod }4 \nonumber \\
\sum_a N_a l^1_a n^2_a n^3_a = \sum_a N_a \tilde{B}_a = 0 \textrm{
mod }4 \nonumber \\
\sum_a N_a n^1_a l^2_a n^3_a = \sum_a N_a \tilde{C}_a = 0 \textrm{
mod }4 \nonumber \\
\sum_a N_a n^1_a n^2_a l^3_a = \sum_a N_a \tilde{D}_a = 0 \textrm{
mod }4
\label{K-charges}
\end{eqnarray}

These constraints turn out to be more clear if the additional
three D5-branes or D9-brane are introduced as ``probes'' \cite{UR}.
These branes wrap cycles along the O6-planes so they satisfy supersymmetry
automatically and form $USp$ groups.  The sum of intersection
numbers between these probe branes and the general D6-branes
should be even (mod 4 in our case) in order to cancel the global
gauge anomaly \cite{Witten}.  For example,
\begin{equation}
\sum_a N_a [\Pi_{D5_1}][\Pi_a]= \sum_a N_a l^1_a n^2_a n^3_a = 0
\textrm{ mod }4
\end{equation}
which is exactly the same as the second equation in (\ref{K-charges}).  Though
we add these extra branes to detect the K-theory charges, they are
still exterior to our original model and do not contribute to the
determined RR-tadpole cancellation configuration.  

\begin{table}[h]
\renewcommand{\arraystretch}{1.25}
\center
\begin{tabular}{|c||l@{}l|l@{}l|}
\hline

D3-brane & $\Pi_{D3}$&$=([b_1])([b_2])([b_3])$ &
$\Pi_{\overline{D3}}$&$=(-[b_1])(-[b_2])(-[b_3])$   \\ \hline

         & $\Pi_{D5_1}$&$=([a_1])([b_2])([b_3])$ &
$\Pi_{\overline{D5}_1}$&$=([a_1])(-[b_2])(-[b_3])$   \\

D5-brane & $\Pi_{D5_2}$&$=([b_1])([a_2])([b_3])$ &
$\Pi_{\overline{D5}_2}$&$=(-[b_1])([a_2])(-[b_3])$   \\

         & $\Pi_{D5_3}$&$=([b_1])([b_2])([a_3])$ &
$\Pi_{\overline{D5}_3}$&$=(-[b_1])(-[b_2])([a_3])$   \\ \hline

         & $\Pi_{D7_1}$&$=([b_1])([a_2])([a_3])$ &
$\Pi_{\overline{D7}_1}$&$=(-[b_1])([a_2])([a_3])$   \\

D7-brane & $\Pi_{D7_2}$&$=([a_1])([b_2])([a_3])$ &
$\Pi_{\overline{D7}_2}$&$=([a_1])(-[b_2])([a_3])$   \\

         & $\Pi_{D7_3}$&$=([a_1])([a_2])([b_3])$ &
$\Pi_{\overline{D7}_3}$&$=([a_1])([a_2])(-[b_3])$   \\ \hline

D9-brane & $\Pi_{D9}$&$=([a_1])([a_2])([a_3])$ &
$\Pi_{\overline{D9}}$&$=([a_1])([a_2])([a_3])$   \\ \hline

\end{tabular}
\caption{Brane pairs of Type IIB theory without B-field and their 
corresponding homology classes of 3-cycles in type
IIA picture.}
\end{table}

\section{Model Building for a Flipped $SU(5)$ GUT}

\subsection{Basic Flipped $SU(5)$ Phenomenology}

In a flipped $SU(5)\times U(1)_X$ \cite{BARR,IJDJ} unified model,
the electric charge generator $Q$ is only partially embedded in
$SU(5)$, {\it i.e.}, $Q = T_3 - \frac{1}{5}Y' +
\frac{2}{5}\tilde{Y}$, where $Y'$ is the $U(1)$ internal $SU(5)$
and $\tilde{Y}$ is the external $U(1)_X$ factor.  Essentially,
this means that the photon is \lq shared\rq \ between $SU(5)$ and
$U(1)_X$. The Standard Model (SM) plus right handed neutrino
states reside within the representations $\bar{\bf{5}}$,
$\bf{10}$, and $\bf{1}$ of $SU(5)$, which are collectively
equivalent to a spinor $\bf{16}$ of $SO(10)$.  The quark and
lepton assignments are flipped by $u^c_L$ $\leftrightarrow$
$d^c_L$ and $\nu^c_L$ $\leftrightarrow$ $e^c_L$ relative to a
conventional $SU(5)$ GUT embedding:
\begin{equation}
\bar{f}_{\bf{\bar{5},-\frac{3}{2}}}= \left( \begin{array}{c}
              u^c_1 \\ u^c_2 \\ u^c_3 \\ e \\ \nu_e
                    \end{array} \right) _L ; \;\;\;
F_{\bf{10,\frac{1}{2}}}= \left( \left( \begin{array}{c}
              u \\ d \end{array} \right) _L  d^c_L \;\; \nu^c_L
\right)
              ; \;\;\;
l_{\bf{1,\frac{5}{2}}}=e^c_L
\end{equation}
In particular this results in  the $\bf{10}$ containing a neutral component with the
quantum numbers of $\nu^c_L$.  We can break spontaneously the GUT
symmetry by using a $\bf{10}$ and $\overline{\bf{10}}$ of
superheavy Higgs where the neutral components provide a large
vacuum expectation value, $\left\langle \nu^c_H \right\rangle$=
$\left\langle \bar{\nu}^c_H \right\rangle$,
\begin{equation}
H_{\bf{10,\frac{1}{2}}}=\left\{Q_H,\;d^c_H,\;\nu^c_H \right\}; \;\;\;
\bar{H}_{\bf{\overline{10},-\frac{1}{2}}}=\left\{Q_{\bar{H}},\;d^c_{\bar{H}},\;\nu^c_{\bar{H}}
\right\}.
\end{equation}
The electroweak spontaneous breaking is generated by the Higgs
doublets $H_2 $ and $ \bar{H}_{\bar{2}} $
\begin{equation}
h_{\bf{5,-1}}=\left\{ H_2,H_3 \right\}; \;\;\;
\bar{h}_{\bf{\bar{5},1}}=\left\{
\bar{H}_{\bar{2}},\bar{H}_{\bar{3}} \right\}
\end{equation}
Flipped $SU(5)$ model building has two very nice features which
are generally not found in typical unified models: (i) a natural
solution to the doublet ($H_2$)-triplet($H_3$) splitting problem
of the electroweak Higgs pentaplets $h,\bar{h}$ through the
trilinear coupling of the Higgs fields: $H_{\bf{10}} \cdot
H_{\bf{10}} \cdot h_{\bf{5}} \rightarrow \left\langle \nu^c_H
\right\rangle d^c_H H_3$, and (ii) an automatic see-saw mechanism
that provide heavy right-handed neutrino mass through the coupling
to singlet fields $\phi$, $F_{\bf{10}} \cdot {\bar
H}_{\overline{\bf{10}}} \cdot \phi \rightarrow \left\langle
\nu^c_{\bar{H}}\right\rangle \nu^c \phi$.

The generic superpotential $W$ for a flipped $SU(5)$ model will be of the form :
\begin{equation}
\lambda_1 FFh+\lambda_2 F\bar{f}\bar{h}+\lambda_3 \bar{f}l^c h+ \lambda_4 F\bar{H}\phi +\lambda_5 HHh+\lambda_6 \bar{H}\bar{H}\bar{h}+ \cdots\in W
\end{equation}
the first three terms provide masses for the quarks and leptons,
the fourth is responsible for the heavy right-handed neutrino mass
and the last two terms are responsible for the doublet-triplet
splitting mechanism \cite{IJDJ}.
\subsection{Model Building}

This model is similar to the one given in \cite{nkwmm}, however  the
K-theory constraints are satisfied.  
In this Letter, we present an example with 6+1 stacks of branes.
The first stack has the same
set of wrapping numbers as in our previous model \cite{nkwmm}.  
We also have a stack with $N^{(\Omega R)}=8$ filler branes which 
give rise to a $USp(8)$ gauge group. 
The gauge symmetry of the (6+1)-stack model, whose wrapping numbers are 
presented in Table 3, is $U(5)\times U(1)^5\times USp(8)$, and the 
structure parameters of the wrapping space are 
\begin{equation}
x_A=1,\;\;\;x_B=2,\;\;\;x_C=10,\;\;\;x_D=1
\end{equation}
The intersection numbers are listed 
in Table 4, and the resulting spectrum in Table 5.


The singlet (under the $SU(5)$ symmetry) representation $e^c_L$, now comes from the bi-fundamentals, namely from the intersection $(cf)$ and we
choose the $\textbf{5}$ and $\overline{\textbf{5}}$ Higgs pentaplets from a
non-chiral interesection $(ab^{\prime})$.  There is less 
exotic matter in this model, though we still have
two copies of $\overline{\textbf{15}}$ which is  unavoidable since we need
$\overline{\textbf{10}}$ in the Higgs sector.
Matter charged under both the $SU(5)\times U(1)_X$
and $USp(8)$ gauge symmetries is also present, as is evident from
Table 4.

The $U(1)_X$ is
\begin{equation}
U(1)_X=\frac{1}{2} \left( U(1)_a-5U(1)_b+5U(1)_c+5U(1)_d-5U(1)_e-5U(1)_f 
\right)
\end{equation}
while the other anomaly-free and massless combinations $U(1)_Y$ is
\begin{equation}
U(1)_Y = U(1)_b-U(1)_c+U(1)_d-U(1)_e
\end{equation}

The remaining four global $U(1)$s from the Green-Schwarz mechanism
are given respectively by
\begin{eqnarray}
 U(1)_1 & = & -10U(1)_a+2U(1)_b+2U(1)_c-2U(1)_d-2U(1)_e-2U(1)_f  \nonumber\\
 U(1)_2 & = & -2U(1)_b-2U(1)_c   \nonumber\\
 U(1)_3 & = & 8U(1)_b+8U(1)_c+4U(1)_d+4U(1)_e   \nonumber\\
 U(1)_4 & = & 20U(1)_a+8U(1)_b+8U(1)_c+4U(1)_f
\end{eqnarray}

\begin{table}[h]
\begin{tabular}{|c|c||ccc||c|c|c|c||c|c|c|c|}
\hline

stack& $N_a$ & ($n_1$, $l_1$) & ($n_2$, $l_2$) & ($n_3$, $l_3$) & $A$ &
$B$ & $C$ & $D$ & $\tilde{A}$ & $\tilde{B}$ & $\tilde{C}$ &
$\tilde{D}$  \\ \hline \hline

$a$ & $N=10$ & ( 0,-1) &(-1,-1) & (-1,-2) & 0 & 0 & -2 & -1 & 2
& -1 & 0 & 0 \\ \hline

 $b$ & $N=2$ & (-1,-1) & (-1, 1) & ( 1, 4) & -1 & -4 & 4 & -1 & 4 &
1 & -1 & 4  \\ \hline

 $c$ & $N=2$ & (-1,-1) & (-1, 1) & ( 1, 4) & -1 & -4 & 4 & -1 & 4 &
1 & -1 & 4  \\ \hline

 $d$ & $N=2$ & (-1, 1) & ( 1, 0) & (-1,-2) & -1 & 0 & -2 & 0 & 0 & -1
& 0 & 2  \\ \hline

 $e$ & $N=2$ & (-1, 1) & ( 1, 0) & (-1,-2) & -1 & 0 & -2 & 0 & 0 & -1
& 0 & 2  \\ \hline

 $f$ & $N=2$ & ( 0, 1) & ( 1, 1) & (-1,-2) & 0 & 0 & -2 & -1 & 2 & -1
& 0 & 0  \\ \hline

filler & $N^{(\Omega R)}=8$ & ( 1, 0) & ( 1, 0) & ( 1, 0) & -1 & 0
& 0 & 0 & 0 & 0 & 0 & 0
\\ \hline
\end{tabular}
\caption{Wrapping numbers and their consistent parameters.}
\end{table}

\begin{table}[h]
\begin{center}
\footnotesize
\begin{tabular}{|c|c||c|c||c|c|c|c|c|c|c|c|c|c||c|}
\hline

stk & $N$ & A & S & $b$ & $b'$ & $c$ & $c'$ & $d$ & $d'$ & $e$ &
$e'$ & $f$ & $f'$ & f1  \\ \hline \hline

$a$ & 10 & 2 & -2 & -4 & 0(6) & -4 & 0(6) & 0(1) & 4 & 0(1) &
4 & 0(0) & 0(8) & 2  \\ \hline

$b$ & 2 & 32 & 0 & - & - & 0(0) & 32 & 4 & 0(6) & 4 & 0(6) &
4 & 0(6) & 4  \\ \hline

$c$ & 2 & 32 & 0 & - & - & - & - & 4 & 0(6) & 4 & 0(6) & 4 &
0(6) & 4  \\ \hline

$d$ & 2 & 2 & -2 & - & - & - & - & - & - & 0(0) & 0(8) &
0(1) & 4 & 0  \\ \hline

$e$ & 2 & 2 & -2 & - & - & - & - & - & - & - & - & 0(1) & 4 & 0  \\ \hline

$f$ & 2 & 2 & -2 & - & - & - & - & - & - & - & - & - & - & 2  \\ \hline

\end{tabular}
\caption{List of intersection numbers.  The number in parenthesis
indicates the multiplicity of non-chiral pairs.}
\end{center}
\end{table}

\begin{table}[h]
\begin{center}
\footnotesize
\begin{tabular}{|c||@{}c@{}||@{}c@{}|@{}c@{}|@{}c@{}|@{}c@{}|@{}c@{}|@{}c@{}||@{}c@{}||@{}c@{}|
@{}c@{}|@{}c@{}|@{}c@{}||@{}c@{}|} \hline

 Rep. & Multi. &$U(1)_a$&$U(1)_b$&$U(1)_c$& $U(1)_d$
& $U(1)_e$ & $U(1)_f$ & $2U(1)_X$ & $U(1)_1$ & $U(1)_2$ &
$U(1)_3$ & $U(1)_4$ & $U(1)_Y$  \\
\hline \hline

$(10,1)$ & 3 & 2 & 0 & 0 & 0 & 0 & 0 & 2 & -20 & 0 & 0 & 40 & 0 \\

$(\bar{5}_a ,1_b)$ & 3 & -1 & 1 & 0 & 0  & 0 & 0 & -6 & 12 & -2 & 8
& -12 & 1  \\

$(1_c ,\bar{1}_f)$ & 3 & 0 & 0 & 1 & 0 & 0 & -1 & 10 & 4 & -2 & 8 & 4 & -1
\\ \hline \hline

$(10,1)$ & 1 & 2 & 0 & 0 & 0 & 0 & 0 & 2 & -20 & 0 & 0 & 40 & 0
\\

$(\overline{10},1)$ & 1 & -2 & 0 & 0 & 0 & 0 & 0 & -2 & 20 & 0 & 0
& -40 & 0  \\
\hline

$(5_a,1_b)^\star$ & 1 & 1 & 1 & 0 & 0 & 0 & 0 & -4 & -8 & -2 & 8 &
28 & 1   \\

$(\bar{5}_a ,\bar{1}_b)^\star$ & 1 & -1 & -1 & 0 & 0 & 0 & 0 & 4 &
8 & 2 & -8 & -28 & -1
\\ \hline

$(1_b ,1_c)$ & 4 & 0 & 1 & 1 & 0 & 0 & 0 & 0 & 4 & -4 & 16 & 16 & 0
\\  \hline \hline

$(\overline{15},1)$ & 2 & -2 & 0 & 0 & 0 & 0 & 0 & -2 & 20 & 0 & 0
& -40 & 0   \\

$(\overline{10},1)$ & 1 & -2 & 0 & 0 & 0 & 0 & 0 & -2 & 20 & 0 & 0
& -40 & 0   \\
\hline

$(\bar{5}_a ,1_b)$ & 1 & -1 & 1 & 0 & 0 & 0 & 0 & -6 & 12 & -2 & 8
& -12 & 1   \\

$(\bar{5}_a ,1_c)$ & 4 & -1 & 0 & 1 & 0 & 0 & 0 & 4 & 12 & -2 & 8
& -12 & -1   \\

$(5_a ,1_d)$ & 4 & 1 & 0 & 0 & 1 & 0 & 0 & 6 & -12 & 0 & 4 & 20 & 1   \\

$(5_a ,1_e)$ & 4 & 1 & 0 & 0 & 0 & 1 & 0 & -4 & -12 & 0 & 4 & 20 & -1
 \\ \hline

$(1_b ,1_c)$ & 28 & 0 & 1 & 1 & 0 & 0 & 0 & 0 & 4 & -4 & 16 & 16 & 0   \\

$(1_b ,\bar{1}_d)$ & 4 & 0 & 1 & 0 & -1 & 0 & 0 & -10 & 4 & -2 & 4
& 8 & 0   \\

$(1_b ,\bar{1}_e)$ & 4 & 0 & 1 & 0 & 0 & -1 & 0 & 0 & 4 & -2 & 4
& 8 & 2   \\

$(1_b ,\bar{1}_f)$ & 4 & 0 & 1 & 0 & 0 & 0 & -1 & 0 & 4 & -2 & 8
& 4 & 1   \\

$(1_c ,\bar{1}_d)$ & 4 & 0 & 0 & 1 & -1 & 0 & 0 & 0 & 4 & -2 & 4
& 8 & -2   \\

$(1_c ,\bar{1}_e)$ & 4 & 0 & 0 & 1 & 0 & -1 & 0 & 10 & 4 & -2 & 4
& 8 & 0   \\

$(1_c ,\bar{1}_f)$ & 1 & 0 & 0 & 1 & 0 & 0 & -1 & 10 & 4 & -2 & 8
& 4 & -1   \\

$(1d ,1_f)$ & 4 & 0 & 0 & 0 & 1 & 0 & 1 & 0 & -4 & 0 & 4 & 4 & 1  \\

$(1e ,1_f)$ & 4 & 0 & 0 & 0 & 0 & 1 & 1 & -10 & -4 & 0 & 4 & 4 & -1  \\

$(\bar{1},\bar{1})$ & 2 & 0 & 0 & 0 & -2 & 0 & 0 & -10 & 4 & 0 & -8 & 0 & -2   \\

$(\bar{1},\bar{1})$ & 2 & 0 & 0 & 0 & 0 & -2 & 0 & 10 & 4 & 0 & -8 & 0 & 2   \\

$(\bar{1},\bar{1})$ & 2 & 0 & 0 & 0 & 0 & 0 & -2 & 10 & 4 & 0 & 0 & -8 & 0   \\
\hline

$(5_a ,1_b)^\star$ & 5 & 1 & 1 & 0 & 0 & 0 & 0 & -4 & -8 & -2 & 8 & 28 & 1   \\

$(\bar{5}_a ,\bar{1}_b)^\star$ & 5 & -1 & -1 & 0 & 0 & 0 & 0 & 4 &
8 & 2 & -8 & -28 & -1  \\
\hline

\multicolumn{14}{|c|}{Additional non-chiral Matter}\\ \hline
\multicolumn{14}{|c|}{USp(8) Matter}\\ \hline\hline
\end{tabular}
\caption{The spectrum of $U(5)\times U(1)^5\times USp(8)$, or
$SU(5)\times U(1)_X\times U(1)_Y\times USp(8)$, with
the four global $U(1)$s from the Green-Schwarz mechanism.  The
$\star'd$ representations stem from vector-like non-chiral pairs.}
\end{center}
\end{table}

\clearpage

\section{Conclusions}

In this Letter we have constructed a particular $N=1$
supersymmetric three-family model whose gauge symmetry includes
$SU(5)\times U(1)_X$, from type IIA orientifolds on $
T^6/(\Z_2\times \Z_2)$ with D6-branes intersecting at general
angles. The model satisfies all the consistency requirements of string 
theory and in addition the constraints arising from the K-theory 
interpretation of D-branes.

The spectrum contains a complete grand unified theory and
electroweak Higgs sector, the latter however, in a non-minimal 
number of copies. In addition, it contains extra 
matter both in bi-fundamental and vector-like representations as
well as two copies of matter in the symmetric representation of
$SU(5)$. Chiral matter charged under both the $SU(5)\times U(1)_X$
and $USp(8)$ gauge symmetries is also present, as is evident from
Table 4. Furthermore, three adjoint $(N=1)$ chiral multiplets are
provided from the $aa$ sector \cite{Cvetic}.

The global symmetries, that arise after the G-S anomaly
cancellation mechanism, forbid some of the Yukawa couplings
required for mass generation, for instance terms like $FFh$.
However, by the same token the term $HHh$ is also forbidden. We
note that such a term is essential for the doublet-triplet
splitting solution mechanism in flipped $SU(5)$.  
On the other hand, the global $U(1)$ symmetries do not forbid 
the coupling $F\bar{f}\bar{h}$, responsible  for the up-type ($t,c,u$) 
quark mass terms.
Neutrino Yukawas ($F\bar{H}\phi$), and the coupling $\bar{f}l^c h$
are also absent at the trilinear level, due to the global $U(1)$ symmetries.
Nevertheless, it
should not escape our notice that while these global $U(1)$
symmetries are exact to all orders in perturbation theory, they
can be broken explicitly by non-perturbative instanton
effects \cite{kachru}, thus providing us with the possibility of
recovering the appropriate superpotential couplings. Another
interesting approach toward generating these absent Yukawa
couplings may entail the introduction of type IIB flux
compactifications \cite{cjan05}. This exceeds the scope of our
current Letter, but shall be further investigated in an upcoming
publication.

An additional important avenue for future research is 
avoidance of  the chiral supermultiplets in the adjoint 
representation from the $aa$ sector.
These multiplets are associated with the moduli space of deformations 
of special Lagrangian submanifolds and their number is equal to the first 
Betti number of the 3-cycle that the stack $a$ of D6-branes wrap \cite{MCLEAN}.
Mechanisms for blocking these problematic particles for phenomenology
include constructing 3-cycles of the wrapping space with zero first Betti-number (such as $S^3$ or $\R P^3$ \footnote{Such Lagrangian manifolds are additionally interesting because they are stable and volume 
minimizing under 
Hamiltonian deformations \cite{Oh}.}) or other rigid 3-cycles, as  has been recently discussed in \cite{MIR1}.
Alternatively, it may be possible to simply give masses to these particles 
at very high energies.
These issues however, are also beyond the scope of this Letter and shall 
be investigated in a future publication.

\section{Acknowledgments}

The work of G. V. K. and D. V. N. is supported by DOE grant 
DE-FG03-95-Er-40917.

\end{document}